\tolerance=10000
\input phyzzx

\REF\cst{C.M. Hull, Nucl.Phys. {\bf B583} (2000) 237, 
hep-th/0004195.}
\REF\jhs{J. H. Schwarz, Talk presented at the Ninth Marcel Grossmann Meeting
(MG9), hep-th/0008009.}
\REF\roz{M. Rozali, Phys. Lett. {\bf  B400} (1997) 260,
hep-th/9702136.}
\REF\brs{M. Berkooz, M. Rozali, and N. Seiberg,
Phys. Lett. {\bf B408} (1997) 105,
hep-th/9704089.}
\REF\note{N.Seiberg, hep-th/9705117.}
\REF\hen{X. Bekaert, M. Henneaux and A. Sevrin, Phys. Lett. {\bf
B468} (1999) 228; hep-th/9909094.}
\REF\strom{A. Strominger, Phys. Lett. {\bf B383} (1996) 44, hep-th/9512059;
P.K. Townsend, Phys. Lett. {\bf B373} (1996) 68, hep-th/9512062.}
 \REF\Cremmer{E. Cremmer, in {\it Supergravity and Superspace},
S.W.
Hawking and
M. Ro\v cek, C.U.P.
Cambridge,  1981.}
\REF\HT{C.M. Hull and P.K. Townsend, Nucl. Phys. {\bf B438} (1995)
109;  hep-th/9410167.}
\REF\Witten{E. Witten, Nucl. Phys. {\bf B443} (1995)  85,
hep-th/9503124.}
\REF\HStr{C.M. Hull, Nucl. Phys. {\bf B468} (1996) 113,
hep-th/9512181.}
\REF\notivarg{
S.~Deser, P.~K.~Townsend and W.~Siegel,
Nucl.\ Phys.\  {\bf B184}, 333 (1981).}

%

%

\def \aa {\alpha}
\def \bb {\beta}

\def \dd {\delta}
\def \ee {\epsilon}

\def \kk {\kappa}

\def \mm {\mu}
\def \nn {\nu}

\def \zz {\zeta}

\def \ww{\omega}

 \def \ggg {\Gamma}

\def \www{\Omega}

\def \2 {{1 \over 2}}
\def \3 {{1 \over 3}}
\def \4 {{1 \over 4}}
\def \5 {{1 \over 5}}
\def \6 {{1 \over 6}}
\def \7 {{1 \over 7}}
\def \8 {{1 \over 8}}
\def \9 {{1 \over 9}}
\def \00 { \infty}

\def\++ {{(+)}}
\def \- {{(-)}}
\def\+-{{(\pm)}}

\def\ek {\eqn\abc$$}

\def \pa {\partial}


 \def\unit{\hbox to 3.3pt{\hskip1.3pt \vrule height 7pt width .4pt
\hskip.7pt
\vrule height 7.85pt width .4pt \kern-2.4pt
\hrulefill \kern-3pt
\raise 4pt\hbox{\char'40}}}

\def\nup#1({Nucl.\ Phys.\  {\bf B#1}\ (}

\def \qq {\qquad}


\Pubnum{ \vbox{ \hbox {QMW-00-11} \hbox{hep-th/0011171}} }
\pubtype{}
\date{November, 2000}

\titlepage

\title {\bf  
  Conformal Non-Geometric   Gravity in Six Dimensions
 and M-Theory above the Planck 
Energy }

\author{C.M. Hull}
\address{Physics Department,
Queen Mary and Westfield College,
\break
Mile End Road, London E1 4NS, U.K.}
\andaddress{Centre Emil Borel,
Institut Henri Poincar\' e,
\break
11 Rue Pierre et Marie Curie,
\break
75231 Paris Cedex 05, France.}
\vskip 0.5cm

\abstract {
The proposal that a strong coupling limit 
of the five-dimensional type II string theory (M-theory  compactified on 
a 6-torus) in which the Planck length becomes infinite could
give a six-dimensional superconformal  phase of M-theory 
is reviewed.
This limit exists for the free theory, giving a 6-dimensional theory with (4,0) supersymmetry compactified on a
circle whose radius gives the 5-dimensional Planck length. 
The free 6-dimensional theory has a fourth rank tensor gauge field with the symmetries of the Riemann
tensor instead of a symmetric tensor gauge field, but its dimensional
reduction gives conventional linearised gravity in five dimensions.
The possibility of an interacting form of this theory existing and the
consequences it
would have  
for the geometric picture of gravity    are discussed. 
\break
{\it Talk given at the Second Gursey Memorial Conference, Istanbul
}

\endpage

Gravitational theories are characterised by a length scale $l$ 
which determines the strength of gravitational coupling and also the 
scale at which quantum gravity effects are expected to become important; 
in M-theory this is 
the $D=11$ Planck length. In recent years there has been great success 
in describing the strong coupling behaviour of a variety of gauge and 
string theories in terms of dual theories. Here, a recent proposal 
[\cst] (reviewed in [\jhs])
to 
extend this to gravitational theories will
 be described.
 The idea is to consider a suitable limit in which $l\to \infty$ 
 so that   masses become zero and a phase of the theory with 
 a vast amount of unbroken gauge symmetry is obtained.
 The specific model is M-theory compactified on $T^{6}$ to $D=5$, with  a
 low-energy effective description in terms of
 $D=5,N=8$ supergravity.
 The proposal is that
 the limit $l\to \infty$ is a kind of  decompactification limit 
 to a six-dimensional theory.
 This theory is conformally invariant, with no length scales, and 
 the scale $l$ arises as the 
 compactification scale: 
  compactification  from $D=6$ on a circle of radius $R$ 
 gives a theory with   $D=5$ Planck scale $l=R$.
 If such a superconformal phase of M-theory exists, then 
 it would have many consequences for our understanding of M-theory and 
 gravitational physics.
 The treatment here will follow [\cst], finishing with some 
 speculations which will be discussed more fully elsewhere.

The aim is then to find a gravitational analogue
of some of the weak-strong coupling dualities that
have been found in field theories and string theories.
At first sight, it would seem that such  dual theories of gravity
would be
unlikely to exist, but the close relations between gravity and
gauge
theories and the implications such a dual
description could have for  quantum
gravity suggest that it could be worthwhile to investigate this
possibility.

A useful example to try and generalise is
 5-dimensional maximally supersymmetric Yang-Mills
theory. It is non-renormalisable and so
new physics should emerge at short distances.
 It will be supposed that the $D=5$
super-Yang-Mills multiplet arises as
part of a consistent theory that can be extrapolated to high energies;
for example, it could arise as part of
the 5-dimensional heterotic string, or  on a stack of D4-branes
in the IIA string, and in both these cases there would be in
addition
supergravity fields and an infinite number of massive fields in
the
full theory.
The theory has   BPS   0-branes given
by
lifting self-dual Yang-Mills instantons in 4
Euclidean dimensions to soliton world-lines in 4+1 dimensions.
These
  have mass
$M \propto {\vert n\vert \over g_{YM}^2}
$
where $n$ is the instanton number, so that these become light as
the dimensionful Yang-Mills coupling
$g_{YM}$ becomes large.
In [\roz] it was proposed that these should be 
interpreted as Kaluza-Klein modes for a $D=6$ theory compactified on 
a circle of radius $R=g_{YM}^2$.
In the strong coupling limit an extra dimension opens up to give
a 6-dimensional (2,0) supersymmetric
theory
in which the gauge field is replaced by a 2-form gauge field $B_{MN}$ with
self-dual field strength,
and the 5 scalar fields are all promoted to scalar
fields in 6 dimensions [\roz-\note].
The $D=6$ theory is believed to be
a non-trivial superconformally invariant quantum theory [\note]
and the $D=5$ gauge coupling 
arises as the radius of the compactification circle, $g_{YM}^2= R$.
The relationship between the $D=5 $ and $D=6$ theories is
straightforward to establish for the free case  in which
the Yang-Mills gauge group is abelian, but in the interacting
theory
the 6-dimensional origin of the $D=5$ non-abelian interactions
is mysterious; there are certainly no
local covariant interactions
that can be written down that give Yang-Mills interactions when
dimensionally
reduced [\hen].
Nonetheless, the fact that these $D=5$
and $D=6$ theories arise
as the world volume theories of D4 and M5 branes respectively   [\strom]
gives
strong support for the existence of such a 6-dimensional origin
for the gauge
interactions. 

The W-bosons and magnetic strings in $D=5$ arise from self-dual 
strings in $D=6$. At the origin of moduli 
space the W-bosons become massless and the
  tensions of the self-dual 
strings in $D=6$ must also become zero.
The nature of the theory at such points is unclear.  For example,
one
guess might be that it could be   described by some kind of
string field theory, but it was argued in [\note] that such a
description
would  over-count the degrees of
freedom.
Nonetheless, given that mysterious  interactions
with no conventional field theory formulation
arise in the M5-brane
world-volume theory, it is natural to ask whether
similar  unconventional interactions could arise elsewhere;
given that they arise in one corner of M-theory, it seems
reasonable
to expect that there are other corners in which such phenomena
occur.

The theory that will be considered here
is five-dimensional $N=8$ supergravity (ungauged),
which has a global $E_6$ symmetry and a local $Sp(4)=USp(8)$
symmetry [\Cremmer].
It is non-renormalisable, and will be regarded as arising as a
massless sector
of some consistent theory, such as
M-theory compactified on a 6-torus, in which the global $E_6$
symmetry is broken to a discrete U-duality subgroup [\HT].
The massless bosonic fields consist of a graviton,
27 abelian vector fields and 42 scalars.
The action
is
$$S=\int d^{5}x \sqrt{-g}\left( {1\over l^{3}}R-
{1\over 4l^{}}F^{2}+\ldots \right)
\eqn\act$$
where $l=\kk ^{2/3}$ is the 5-dimensional Planck length.
If this does have a  dual at strong gravitational coupling, i.e. a
limit as
$l\to \infty$, then
it is natural to look
first for a theory with 32 supersymmetries and
with $E_{6 }\times Sp(4)$ symmetry, as
the  simplest possibility would be if these symmetries
survived at strong coupling.
The arguments used in [\cst] are similar to those used in
  [\HT,\Witten,\HStr] for extrapolating to large values of
dimensionless
  string couplings and moduli, and
it will be assumed that all symmetries are preserved
and BPS states are protected and survive as the coupling $l$ is
increased.
The symmetries then impose stringent constraints on what can
happen,
and
one can then seek checks on the predicted strong-coupling  dual.

Comparison with the $D=5$ gauge theory case suggests seeking
a 6-dimensional theory
compactified on a circle in which the 27 $D=5$ vector fields
are replaced
with
27 self-dual 2-form gauge fields in $D=6$.
Indeed, the decomposition of $ N=8$ supergravity into $N=4$
multiplets
includes five $N=4$ vector multiplets with coupling
constant $g_{YM}^{2}=l$, and the strong coupling limit of each should give a
(2,0)
6-dimensional
tensor multiplet [\cst].
However, the 27 vector fields of the $N=8$ theory fit into an
irreducible representation of $E_{6}$, and so if $E_{6}$ symmetry
is
to be maintained and if    some of the vector fields become
self-dual 2-form gauge fields in 6 dimensions, then all of them
should.
Furthermore, all the fields fit into an irreducible multiplet of
the
$N=8$ supersymmetry
algebra, so that
if the 32 supersymmetries survive at strong coupling and some of
the
fields become 6-dimensional, then the whole theory should become
6-dimensional.
As in the gauge theory case, all the scalar fields should survive
at
strong coupling and so should lift to
   42 scalars in 6 dimensions.
If the $D=6$ theory were conformally invariant, then
the $D=5$ gravitational coupling $l$ could
arise from the radius of the circle $R$.

This then suggests that the strong coupling dual field theory
should
be
 a superconformal theory in six dimensions
with 32 ordinary supersymmetries (and   a further 32 conformal
supersymmetries),  and should have
42 scalars and 27 self-dual 2-form gauge fields. The unique
supergravity theory in 6 dimensions with 32 supersymmetries
fails on every one of these requirements with the obvious
exception
  of having
32 supersymmetries.
Remarkably, such a conformal supermultiplet in six dimensions
with 32 supersymmetries, $Sp(4)$ R-symmetry, 42 scalars and 27
self-dual 2-form gauge
fields {\it
does } exist and
so is an immediate candidate for a strong coupling
dual. 

The multiplet has (4,0) supersymmetry in $D=6$ and was studied in 
[\cst].
Instead of a graviton, it has an exotic
fourth-rank tensor gauge field
 $C_{MN\, PQ}
$ with the algebraic properties of the Riemann tensor
and the gauge symmetry
$$ \dd C_{MN\, PQ} = \pa _{[M} \chi _{N]PQ} +\pa _{[P} \chi
_{Q]MN}-2
\pa _{[M} \chi _{NPQ]}
\eqn\delcis$$
with parameter $\chi _{MPQ}=-\chi _{MQP}$.
(Similar gauge fields were considered in $D=4$ in [\notivarg].)
The invariant field strength is
$$G_{MNP\, QRS} ={1\over 36}(\pa_{M}\pa_{S}C_{NP\, RS}+\ldots)=
\pa _{[M} C_{NP]\,  [QR,S]}
\eqn\gis$$
and in the (4,0) multiplet it satisfies the self-duality constraint
$$G_{MNP\, QRS}
= \6 \epsilon _{MNPTUV}
G^{TUV}{}_{ QRS}
\eqn\gdu$$
or $G=*G$.
In addition, there are 27 2-form gauge-fields with self-dual field 
strengths, 42 scalars, 48 symplectic Majorana-Weyl fermions and, 
instead of gravitini, 8 spinor-valued 2-forms $\Psi _{MN }^\aa
$ which satisfy a symplectic Majorana-Weyl
constraint and have self-dual field strengths.
The fermionic gauge symmetry  is of the form
$$\dd \Psi _{MN }^a =
\pa_{[M}\varepsilon_{N]}^a 
\eqn\delyis$$
with parameter a spinor-vector $\varepsilon_{N}^{\aa a}$. 
 The free theory based on this
multiplet
is a superconformally invariant theory, with conformal
supergroup $OSp^*(8/8)$  [\cst]. This has bosonic subgroup
$USp(8)\times SO^*(8)=Sp(4)\times SO(6,2)$ and 64 fermionic
generators, consisting of the 32 supersymmetries of the (4,0) 
superalgebra and 32 conformal supersymmetries.

It is remarkable that in going from $D=5$ to $D=6$, the 
vector gauge fields $A_{\mm}$ are lifted to 
2-forms $B_{MN}$, the gravitini $\psi_{\mm}$ are   lifted to 
spinor-valued
2-forms $\Psi_{MN}$ and the graviton $h_{\mm\nn}$ is lifted to the gauge 
field  $C_{MN\, PQ}
$, with these $D=6$ gauge-fields all satisfying self-duality 
constraints.
Electrically charged 0-branes and magentic strings in $D=5$   
lift to BPS self-dual strings in $D=6$.
In [\cst] it was shown that the dimensional reduction of the free (4,0) theory on 
a circle indeed gives the 
linearised $D=5,N=8$ supergravity theory, with
gravitational coupling (Planck length)   given by the circle 
radius $l=R$.
However, 
 there are no covariant local interactions in $D=6$ for this multiplet
that could give rise to
the $D=5$ supergravity interactions.

There is then a close analogy between the
$D=5,N=4$ Yang-Mills theory and the 
$D=5,N=8$ supergravity.
The linearised versions of these theories both arise from the 
dimensional reduction of a superconformal field 
theory in $D=6$, with the dimensional $D=5$ coupling constant 
arising from the radius of compactification, so that the strong 
coupling limit of the $D=5$ free theories is a decompactification to $D=6$.
For the interacting $D=5$ Yang-Mills theory, there are a number of 
arguments to support the conjecture that its strong coupling limit 
should be an interacting superconformal theory in $D=6$ with (2,0) 
supersymmetry, even though such a theory has not been constructed and 
indeed cannot have a conventional field theory formulation.
This is not without precedent; in $D=2$, many \lq conformal field 
theories'   do not appear to have  any formulation as conventional field 
theories, and in any dimension, the absence of asymptotic particle 
states suggests that these are better formulated not in terms of 
field theory but in terms of correlation functions of operators 
occuring in various representations of the conformal group.
This led to the conjecture of [\cst] that the situation for 
$D=5$ supergravity is similar to that of $D=5$ super-Yang-Mills, 
and that a certain strong coupling limit of the interacting 
supergravity theory should 
give an interacting theory whose free limit is the (4,0) theory in 
$D=6$.
 In this case
there is no analogue of the M5-brane
argument to support this,
although the M5-brane case does set a suggestive
precedent.

 The
 limiting theory should have some magical form of interactions
 which give the non-polynomial supergravity interactions on
 reduction. It could be that these are some non-local or
non-covariant
 self-interactions of the (4,0) multiplet, or it could be that
other
 degrees of freedom might be needed;  one candidate might be
 some form of string field theory.
 However, if a strong coupling limit of the theory does exist that
 meets the requirements
 assumed here, then the limit must be  a (4,0) theory in
six
 dimensions, and this would predict the existence of
M-interactions
 arising from the strong coupling limit of the supergravity
 interactions. Although it has not been possible to prove the
 existence of such a limit, it is remarkable that there is such a
 simple candidate theory for the limit with so many properties in
 common with the (2,0) limit of the $D=5$ gauge theory.
 Conversely, if there is a 6-dimensional phase of M-theory which
has
 (4,0) supersymmetry, then its circle  reduction to $D=5$ must
give an $N=8$
 supersymmetric theory and the scenario described here should
apply.

In the context in which the $D=5$ supergravity is the massless
sector
of M-theory on $T^{6}$, then the $D=6$ superconformal theory would be
a field theory sector of
a new 6-dimensional superconformal  phase of M-theory.
This could tell us a great deal  about M-theory: it would 
be
perhaps the most symmetric phase of M-theory so far found, with a 
huge amount of unbroken gauge symmetry, and would be a
phase that is not well-described by a conventional field theory
at low energies, so that it could give new insights into the degrees 
of freedom of M-theory.
The various strong-coupling limits of the 5-dimensional string
theory
(obtained by compactifying M-theory on $T^{6}$) that were
analysed in [\Witten, \HStr] in which
0-branes become massless
all correspond to limits
of the (4,0) theory
in which some string tensions become infinite and others approach 
zero, and an understanding of such tensionless strings will be an 
important part of understanding the (2,0) and (4,0) theories.

Particularly intriguing are the consequences for gravity.
In $D=5$, gravity is described geometrically in terms of a metric 
$g_{\mm\nn}$, but at strong coupling  it is described instead in terms of the 
gauge field $C_{MNPQ}$ in $D=6$ (at least in the free case).
This suggests the possibility of some new structure which reduces to 
Riemannian geometry 
  but which is more general and is the appropriate language for 
  describing  gravity beyond the Planck scale.
  For example, while $g_{\mm\nn}$ provides a norm for vectors and 
  a notion of length, $C_{MNPQ}$ could provide a norm
  $C_{MNPQ}\ww^{MN}\ww^{PQ}$ for 2-forms $\ww^{MN}$ and hence a 
  notion of area that is not derived from a concept of length.
It may be that the interacting theory is not described in 
conventional $D=6$ spacetime at all, but in some other arena.

There seem to be three main possibilities.
The first is that there is no interacting version of the (4,0) theory, 
  that it only exists as a free theory, and that the limit proposed 
  in [\cst] only exists for the free $D=5$ theory.
The second is that an interacting form of the theory does exist in 6 
spacetime dimensions, with $D=6$ diffeomorphism symmetry.
The absence of a spacetime metric means that 
such a generally covariant theory would be of  an unusual kind.
The third and perhaps the most interesting 
possibility is that the theory that reduces to the interacting 
supergravity in $D=5$ is not a diffeomorphism-invariant theory in 
six spacetime dimensions, but is something more exotic. There are some 
suggestions for this from the linear theory, as will now be discussed.

For gravity in any dimension, the full
non-linear gauge symmetry is
$$\dd g_{\mm\nn}=2\nabla_{(\mm}\xi_{\nn)}
\eqn\dif$$
If the metric is written as
$$g_{\mm\nn}=\bar g_{\mm\nn}+h_{\mm\nn}
\eqn\abc$$
in terms of a fluctuation $h_{\mm\nn}$ about 
  some background metric $\bar g_{\mm\nn}$ (e.g. a flat background  metric)
 then two main types of symmetry emerge. 
 The first consists of
 \lq background reparameterizations'
 $$\dd \bar g_{\mm\nn}= 2\bar \nabla_{(\mm}\xi_{\nn)}
 ,\qq
 \dd
h_{\mm\nn}= {\cal L}_{\xi}h_{\mm\nn}
\eqn\bac$$
where
$\bar \nabla$ is the background covariant derivative 
with connection constructed from $\bar g_{\mm\nn}$, while 
$h_{\mm\nn}$ transforms as a tensor (${\cal L}_{\xi}$ is the Lie 
derivative with respect to the vector field $\xi$),
as do all other covariant fields.
The second is the \lq gauge symmetry'
of the form
 $$\dd \bar g_{\mm\nn}=0, \qq
 \dd
h_{\mm\nn}= 2  \nabla_{(\mm}\zz_{\nn)}
\eqn\gag$$
in which 
$h_{\mm\nn}$  transforms as a gauge field and the background is 
invariant.
There is in addition the standard shift symmetry
under which
$$\dd \bar g_{\mm\nn}= \aa_{\mm\nn}
 ,\qq
 \dd
h_{\mm\nn}=- \aa_{\mm\nn}
\eqn\shift$$
In terms of the 
full metric $g_{\mm\nn}$, there is no shift symmetry and a unique gauge
symmetry \dif; the various types of symmetry \bac,\gag,\shift\
are an artifice of the background split.
The shift symmetry is a signal of background independence and plays an
important role in the interacting theory.

The linearised $D=5$ supergravity theory has both 
  background reparameterization and gauge invariances given by the 
  linearised forms of \bac,\gag\ respectively, and both of these have 
  origins in $D=6$ symmetries of the free (4,0) theory.
  The background reparameterization invariance lifts to the 
  linearised $D=6$ background 
  reparameterization invariance
  $$\dd \bar g _{MN}= 2\pa_{(M}\xi_{N)},\qq \dd C_{MNPQ}={\cal L}_{\xi}C_{MNPQ}
  \ek
  with the transformations leaving the flat background metric
$ \bar g _{MN}$ invariant forming the $D=6$ Poincar\' e group.
The $D=5$ gauge symmetry given by the linearised form of \gag\ arises from
the $D=6$ gauge symmetry
\delcis\
with $\dd \bar g _{MN}= 0$ and the parameters related by 
$\zz^{\mm}=\chi^{55\mm}$.
The $D=6$ theory has no analogue of the shift symmetry, and the 
emergence of that symmetry on reduction to $D=5$ and dualising to 
formulate the theory in terms of a graviton $h_{{\mm\nn}}$ comes as a 
surprise from this viewpoint.

The gravitational interactions of the full supergravity theory in $D=5$ 
are best expressed geometrically in terms of the total metric 
$g_{\mm\nn}$.
If an interacting form of the (4,0) theory exists that reduces to the 
$D=5$ supergravity, it must be of an unusual kind.
One possibility is that there is no background metric of any kind in 
$D=6$, and the full theory is formulated in terms of a total field 
corresponding to $C$, with a spacetime metric emerging only in a 
particular background $C$ field and a particular limit corresponding 
to the free theory limit in $D=5$.

It is not even clear that the interacting theory should be formulated 
in a $D=6$ spacetime. 
In $D=5$, the diffeomorphisms act on the coordinates as
$$\dd x^{\mm}=\xi^{\mm}\ek
In the (4,0) theory, the parameter $\xi^{\mm}$ lifts to a parameter
$\chi^{MNP}$.
If the coordinate transformations were to lift, it could be to something like a 
manifold with coordinates $X^{MNP}$ transforming through 
reparameterisations
$$\dd X^{MNP}=\chi^{MNP}
\ek
with the $D=5$ spacetime arising as a submanifold with
$x^{\mm}=X^{55\mm}$.

Similar considerations apply to the local supersymmetry transformations.
In $D=5$, the local supersymmetry transformations in a supergravity 
background give rise to \lq background 
supersymmetry transformations' with symplectic Majorana spinor  parameters
$\ee ^{\aa a} $ 
(where $\aa$ is a $D=5$  spinor index and $a=1,..,8$ labels the 8 
supersymmetries) in which the 
gravitino fluctuation $\psi_{\mm}^{a}$ transforms without a derivative of
$\ee^{a}$, and 
\lq gauge supersymmetries'  with spinor parameter $\varepsilon 
^{\aa a} $ under which 
$$\delta \psi^{a}_{\mm}=\pa_{\mm}
\varepsilon 
^{ a} +\ldots \eqn\ssy$$
The background symmetries preserving a flat space background
form the $D=5$ super-Poincar\' e group.
In the free theory, the $D=5$ super-Poincar\' e symmetry lifts to part 
of a $D=6$ super-Poincar\' e symmetry with $D=5$ translation parameters 
$\xi^{\mm}$ lifting to $D=6$ ones $\Xi^{M}$ and supersymmetry 
parameters $\ee $ lifting to $D=6$ spinor parameters $\hat \ee$.
The corresponding $D=6$ supersymmetry charges $Q$ and momenta $P$
are generators of the (4,0) super-Poincar\' e algebra
with
$$
\{Q_\alpha^a,Q_\beta^b\} =\, \www^{ab}\big( \Pi_+
\Gamma^MC\big)_{\alpha\beta} P_M
\eqn\fialgss$$
where $\Pi_\pm$ are the chiral projectors
$$
\Pi_{\pm}=\2 (1\pm \ggg ^{7})
\eqn\abc$$
$\aa,\bb$ are $D=6$ spinor indices and $a,b=1,\ldots,8$ are $USp(8)$ 
indices, with $\www^{ab}$ the
$USp(8)$-invariant anti-symmetric tensor.
This is in turn part of the $D=6$ superconformal group  $OSp^*(8/8)$.

The $D=5$ gauge symmetries including those with parameters $\zz 
^{\mm},\varepsilon^{a}$ satisfy a local algebra whose global limit is 
the $D=5$ Poincar\' e algebra, but the $D=6$ origin of this (at least 
in the free theory) is an algebra including the
generators ${\cal Q} 
^{a}_{\aa M}$ of the fermionic symmetries with parameter 
$\varepsilon_{N}^{\aa}$ and the generators ${\cal P}^{MNP}$ of the 
bosonic symmetries with parameter $\chi^{MNP}$.
The global algebra is of the form
$$
\{{\cal Q} _{\alpha N}^a,{\cal Q}  _{\beta P}^b\} =\, \www^{ab}\big( \Pi_+
\Gamma^MC\big)_{\alpha\beta} {\cal P}_{(NP)M}
\eqn\fialgsss$$
In the dimensional reduction,
the $D=5$ superalgebra has charges
$Q _{\alpha }^a ={\cal Q} _{\alpha 5}^a$, $P_{\mm}={\cal P}_{55\mm}$.

Supersymmetry provides a further argument against the possibility of 
a background metric playing any role in an interacting (4,0) theory 
in $D=6$. 
The $D=5$ supergravity can be formulated in an arbitrary supergravity 
background, but these cannot be lifted to $D=6$ (4,0) backgrounds
involving a background metric
as there is no (4,0) multiplet including a metric or graviton.
The absence of a (4,0) supergravity multiplet appears to rule  out the 
possibility of a background metric  and
the standard supersymmetry \fialgss\
playing any role in the $D=6$ 
theory. Indeed, the interacting theory (if it exists) should 
perhaps be a theory based on something like
the algebra \fialgsss\ rather than the
super-Poncar\' e algebra \fialgss.

\refout

\bye